\documentstyle[epsfig]{europhys}
\def\And{{\rm and\ }}

\def\stars{\bigskip\centerline{***}\medskip}

\newif\ifboo \boofalse

\def\Review#1{\boofalse{\it #1},}
\def\Name#1{{\sc #1},}
\def\Vol#1{\ifboo Vol. {\bf #1}\else{\bf #1}\fi}
\def\Year#1{\ifboo #1\else(#1)\fi}
\def\Book#1{\bootrue{\it #1},}
\def\Page#1{\ifboo {\rm p. #1}\else{\rm #1}\fi}
\newcommand{\vetor}[1]{\mbox{\boldmath ${#1}$}}

\begin{document}

\euro{}{}{}{2000}
\Date{5 May  2000}
\shorttitle{R. VICENTE et al.  Error-correcting code on a cactus.}
\title{Error-correcting code on a cactus: a solvable model}
\author{R. Vicente\inst{1}, D. Saad\inst{1} and  Y. Kabashima\inst{2} }
\institute{
\inst{1} The Neural Computing Research Group, Aston University, Birmingham B4 7ET, UK \\
\inst{2} Department  of Computational Intelligence and Systems Science, Tokyo Institute of Technology, Yokohama 2268502, Japan} 
%
%%%    The `\maketitle' macro needs the following macro:    \rec{}{}
%%%    to be left empty.
%
%\rec{24 February 2000}{in final form ...}
%

%

%
\pacs{
\Pacs{89}{90+n}{Other areas of general interest to physicists.}
\Pacs{89}{70$+$c}{Information Theory}
\Pacs{05}{50$+$q}{Lattice theory and statistics; Ising problems}
      }
\maketitle

\begin{abstract}
An exact solution to a family of parity check error-correcting codes
is provided by mapping the problem onto a Husimi cactus.  The solution 
obtained in the thermodynamic limit recovers the replica
symmetric theory results and provides a very good approximation to
finite systems of moderate size.  The probability propagation decoding
algorithm  emerges naturally from the analysis. A phase transition
between  decoding success and failure phases is found to coincide
with an information-theoretic upper bound. The method is employed to
compare  Gallager and MN codes.
\end{abstract}

The theory of error-correcting codes concentrates on the efficient
introduction of redundancy to given messages for protecting the
information content against corruption.  The theoretical foundations
of this area were laid by Shannon's seminal work \cite{shannon} and
have been developing ever since (see \cite{viterbi} and references
therein). One of the main results obtained in this field is the
celebrated {\it channel coding theorem } stating that there exists a
code such that the average message error probability $P_E$, when
maximum likelihood decoding is used, is upper bounded by
$P_E<e^{-M\;E(R)}$, where $M$ is the length of the encoded
transmission and $R=($ message information content $)/M$ is the code
rate. The exponent $E(R)$ is positive for code rates below the {\it
channel capacity}, corresponding to the maximal mutual information
between the received and the transmitted signals, and vanishes above
it. For rates $R$ below the channel capacity, commonly termed {\it
Shannon's bound}, the error probability can be made arbitrarily small.

The channel coding theorem is based on unstructured random codes and
impractical decoders as maximum likelihood \cite{viterbi} or typical
 sets \cite{mackay}. In the last
fifty years several practical methods have been proposed and
implemented, but none has been able to saturate Shannon's bound. In
1963 Gallager \cite{gallager} proposed a coding scheme which involves
sparse linear transformations of binary messages that was forgotten
soon after in part due to the success of convolutional codes
\cite{viterbi} and the computational limitations of the
time. Gallager codes have been recently rediscovered by MacKay and
Neal (MN) that independently proposed a  closely related  code
\cite{mackay}. This almost coincided with the breakthrough discovery
of the high performance turbo codes \cite{berrou}. Variations of
Gallager codes have displayed performance comparable (and sometimes
superior) to turbo codes \cite{davey}, qualifying them as
state-of-the-art codes.

Statistical physics has been applied to the analysis of
error-correcting codes as an alternative to information theory methods
yielding some new interesting directions and suggesting new
high-performance codes \cite{IKS}. Sourlas was the first to relate
error-correcting codes to spin glass models \cite{sourlas}, showing
that the Random Energy Model (REM)\cite{derrida,saakian,DW} can be
thought of as an ideal code capable of saturating Shannon's bound at
vanishing code rates. This work was extended recently to the case of finite
code rates \cite{KS,VKS} and has been further developed for analysing
MN codes of various structures \cite{KMS,MKSV,VSK2}.  All of the
analyses mentioned above as well as the recent turbo codes analysis
\cite{montanari} relied on the replica approach under the
assumption of replica symmetry.  It is also worthwhile mentioning a
different approach, used in the analysis of convolutional codes
\cite{dress}, of employing the transfer-matrix formalism and power
series expansions. However, to date, the only  model that can be analysed exactly  is the REM that corresponds to an impractical coding scheme of  a vanishing code rate.

In this letter we present an {\em exact} analysis to the performance
of Gallager error-correcting codes on a generalisation of Bethe lattices
 known as 
Husimi cactus \cite{rieger}. We solve the model recovering results obtained 
by the replica symmetric theory and finding the noise level that corresponds 
to the phase transition between perfect decoding and a decoding failure
phase, this appears to coincide with existing information-theoretic
upper bounds.  We experimentally show that the solution
 accurately approximates Gallager codes  of moderate size. 
We also show that the probability
propagation (PP) decoding algorithm emerges naturally from this
framework allowing for the analysis of the practical decoding performance.
Finally, we summarise the differences between Gallager and MN codes,
which are somewhat obscure in the information theory literature but
become explicit in this framework.

We will concentrate here on a simple communication model whereby
messages are represented by binary vectors and are communicated
through a Binary Symmetric Channel (BSC) where uncorrelated bit flips
appear with probability $f$. A Gallager code is defined by a binary matrix
 $\mbox{\boldmath $A$}=[\mbox{\boldmath $C_1$}\mid\mbox{\boldmath $C_2$}]$, 
concatenating two very sparse matrices known to both sender and receiver, with
$\mbox{\boldmath $C_2$}$ (of dimensionality $(M-N)\times(M-N)$) being
invertible - the matrix $\mbox{\boldmath $C_1$}$ is of dimensionality
$(M-N)\times N$.

Encoding refers to the production of a $M$ dimensional binary code
word $\mbox{\boldmath $t$}\in\{0,1\}^M$ ($M>N$) from the original
message $\mbox{\boldmath $\xi$}\in\{0,1\}^N$ by $\mbox{\boldmath
$t$}=\mbox{\boldmath $G^T$}\mbox{\boldmath $\xi$}\;\mbox{(mod 2)}$,
where all operations are performed in the field $\{0,1\}$ and are
indicated by $\mbox{(mod 2)}$. The generator matrix is
$\mbox{\boldmath $G$}= [\mbox{\boldmath $I$}\mid\mbox{\boldmath
$C_2^{-1}$}\mbox{\boldmath $C_1$}] \mbox{ (mod 2)}$, where $\vetor{I}$
is the $N\times N$ identity matrix, implying that $\mbox{\boldmath
$A$}\mbox{\boldmath $G^T$} \mbox{ (mod 2)} =0$ and that the first $N$
bits of $\mbox{\boldmath $t$}$ are set to the message $\mbox{\boldmath
$\xi$}$. In {\it regular} Gallager codes the number of non-zero
elements in each row of $\vetor{A}$ is chosen to be exactly $K$.  The
number of elements per column is then $C=(1-R)K$, where the code rate
is $R=N/M$ (for unbiased messages).
The encoded vector $\mbox{\boldmath $t$}$ is then corrupted by noise
represented by the vector $\mbox{\boldmath $\zeta$}\in\{0,1\}^M$ with
components independently drawn from
$P(\zeta)=(1-f)\delta(\zeta)+f\delta(\zeta-1)$.  The received vector
takes the form $\mbox{\boldmath $r$}=\mbox{\boldmath
$G^T$}\mbox{\boldmath $\xi$}+\mbox{\boldmath $\zeta$}\mbox{ (mod 2)}$.

Decoding is carried out by multiplying the received message by the
matrix $\vetor{A}$ to produce the {\it syndrome} vector
$\mbox{\boldmath $z$}=\mbox{\boldmath $A$}\mbox{\boldmath
$r$}=\mbox{\boldmath $A$}\mbox{\boldmath $\zeta$}\mbox{ (mod 2)}$ from
which an estimate $\mbox{\boldmath $\widehat\tau$}$ for the noise
vector can be produced. An estimate for the original message is then
obtained as the first $N$ bits of $\mbox{\boldmath $r$}
+\mbox{\boldmath $\widehat\tau$}\mbox{ (mod 2)}$. The Bayes optimal
estimator (also known as {\it marginal posterior maximiser}, MPM) for
the noise is defined as
$\widehat{\tau}_j=\mbox{argmax}_{\tau_j}P(\tau_j\mid z)$. The
performance of this estimator can be measured by the probability of
bit error $p_b=1-1/M \; \sum_{j=1}^M
\delta[\widehat{\tau}_j;\zeta_j]$, where $\delta[;]$ is Kronecker's
delta. Knowing the matrices $\mbox{\boldmath $C_2$}$ and
$\mbox{\boldmath $C_1$}$, the syndrome vector $\vetor{z}$ and the
noise level $f$ it is possible to apply Bayes' theorem and compute the
posterior probability
\begin{equation}
P(\vetor{\tau}\mid \vetor{z})=\frac{1}{Z}\chi\left[\mbox{\boldmath
$z$}=\vetor{A}\vetor{\tau}\mbox{(mod 2)}\right] P(\vetor{\tau}),
\label{eq:posterior}
\end{equation}
where $\chi[X]$ is an indicator function providing $1$ if $X$ is true
and $0$ otherwise. To compute the MPM one has to compute the marginal
posterior $P(\tau_j \mid \vetor{z})= \sum_{i\neq j} P(\vetor{\tau}\mid
\vetor{z})$, which in general requires ${\cal O}(2^M)$ operations,
thus becoming impractical for long messages.  To solve this problem
one can use the sparseness of $\vetor{A}$ to design algorithms that
require ${\cal O}(M)$ operations to perform the same task. One
of these methods is the probability propagation algorithm (PP), also
known as belief propagation, sum-product algorithm (see
\cite{kschischang}) or generalised distributive
law \cite{aji}.

The connection to statistical physics becomes clear when the field
$\{0,1\}$ is replaced by Ising spins $\{\pm 1\}$ and mod $2$ sums  by products \cite{sourlas}. The syndrome vector acquires the
form of a multi-spin coupling ${\cal J}_\mu=\prod_{j\in{\cal
L}(\mu)}\zeta_j$ where $j=1, \cdots, M$ and $\mu=1, \cdots,
(M-N)$. The $K$ indices of nonzero elements in the row $\mu$ of
$\vetor{A}$ are given by ${\cal L}(\mu)=\{ j_1, \cdots, j_K \}$, and
in a column $l$ are given by ${\cal M}(l)=\{ \mu_1, \cdots, \mu_C \}$.
 
The posterior (\ref{eq:posterior}) can  be written as the Gibbs
distribution \cite{KMS,MKSV}:
\begin{eqnarray}
\label{eq:hamiltonian}
P(\vetor{\tau}\mid {\cal
J})&=&\frac{1}{Z}\lim_{\beta\rightarrow\infty}\mbox{exp}\left[-\beta
{\cal H}_\beta(\vetor{\tau};{\cal J})\right] \\
{\cal H}_\beta(\vetor{\tau};{\cal J})&=&-\sum_{\mu=1}^{M-N}\left({\cal J}_\mu
\prod_{j\in{\cal L}(\mu)} \tau_j - 1\right)
-\frac{F}{\beta}\sum_{j=1}^{M}\tau_j \ . \nonumber
\end{eqnarray}
The external field corresponds to the prior probability over the noise
and has the form $F=\mbox{atanh}(1-2f)$. Note that the Hamiltonian
itself depends on the inverse temperature $\beta$.  The disorder is
trivial and can be gauged as ${\cal J}_\mu \mapsto 1$ by using $\tau_j
\mapsto \tau_j\zeta_j$. The resulting Hamiltonian is a multi-spin
ferromagnet with finite connectivity in a random field $h_j=\beta^{-1}
F\zeta_j$. The decoding process corresponds to finding {\it zero
temperature} local magnetisations
$m_j=\lim_{\beta\rightarrow\infty}\langle \tau_j \rangle_\beta$ and
calculating estimates as $\widehat{\tau}_j=\mbox{sgn}(m_j)$.

In the $\{\pm 1\}$ representation the probability of bit error,
acquires the form
\begin{equation}
p_b=\frac{1}{2}-\;\frac{1}{2M}\sum_{j=1}^{M}\zeta_j\mbox{ sgn}(m_j),
\end{equation} 
connecting the code performance with the computation of local
magnetisations. 

\begin{figure}
\begin{center}
\epsfig{file=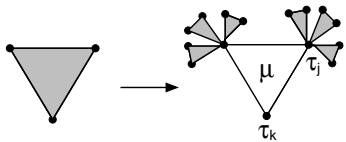,width=45mm}
\caption{First step in the construction of Husimi cactus with $K=3$
and connectivity $C=4$.}
\label{fig1}
\end{center}
\end{figure}

A Husimi cactus with connectivity $C$ is generated starting with a
polygon of $K$ vertices with one Ising spin in each vertex (generation
$0$). All spins in a polygon interact through a single coupling ${\cal
J}_\mu$ and one of them is called the base spin. In figure \ref{fig1}
we show the first step in the construction of a Husimi cactus, in a
generic step the base spins of the $n-1$ generation polygons,
numbering $(C-1)(K-1)$, are attached to $K-1$ vertices of a generation
$n$ polygon. This process is iterated until a maximum generation
$n_{\mbox{\small max}}$ is reached, the graph is then completed by
attaching $C$ uncorrelated branches of $n_{\mbox{\small max}}$
generations at their base spins. In that way each spin inside the
graph is connected to exactly $C$ polygons.  The local magnetisation
at the centre $m_j$ can be obtained by fixing boundary (initial)
conditions in the $0$-th generation and iterating recursion equations
until generation $n_{\mbox{\small max}}$ is reached. Carrying out the
calculation in the thermodynamic limit corresponds to having
$n_{\mbox{\small max}}\sim \ln M$ generations and $M\rightarrow
\infty$.

The Hamiltonian of the model has the form (\ref{eq:hamiltonian}) where
${\cal L}(\mu)$ denotes the polygon $\mu$ of the lattice. Due to the
tree-like structure, local quantities far from the boundary can be 
calculated recursively by specifying boundary conditions.  
The typical decoding performance can
 therefore be computed  exactly without resorting to replica 
calculations \cite{gujrati}.

We adopt the approach presented in \cite{rieger} where recursion
relations for the probability distribution $P_{\mu k}(\tau_k)$ for the
base spin of the polygon $\mu$ is connected to $(C-1)(K-1)$
distributions $P_{\nu j}(\tau_j)$, with $\nu\in{\cal M}(j)\setminus
\mu$ (all polygons linked to $j$ but $\mu$) of polygons in the
previous generation:
\begin{equation}
P_{\mu k}(\tau_k)=\frac{1}{{\cal N}} \mbox{ Tr}_{\{\tau_j\}}
\exp\left[ \beta\left( {\cal J}_\mu \tau_k \prod_{j\in{\cal
L}(\mu)\setminus k}\tau_j-1\right)+ F\tau_k \right]\prod_{\nu\in{\cal
M}(j)\setminus \mu}\prod_{j\in{\cal L}(\mu)\setminus k}P_{\nu
j}(\tau_j),
\label{eq:recursive}
\end{equation}
where the trace is over the spins $\tau_j$ such that $j\in{\cal
L}(\mu)\setminus k$.

The effective field $\widehat{x}_{\nu j}$ on a base spin $j$ due to
neighbours in polygon $\nu$ can be written as :
\begin{equation}
  \exp\left(-2 \widehat{x}_{\nu j} \right)=\mbox{e}^{2F}\frac{ P_{\nu
  j}(-)}{P_{\nu j}(+)},
\label{eq:aux_x}
\end{equation}
Combining (\ref{eq:recursive}) and (\ref{eq:aux_x}) one finds the
recursion relation:
\begin{equation}
\exp\left(-2 \widehat{x}_{\mu k}\right)=\frac{\mbox{ Tr}_{\{\tau_j\}}
\exp\left[- \beta {\cal J}_\mu \prod_{j\in{\cal L}(\mu)\setminus
k}\tau_j + \sum_{j\in{\cal L}(\mu)\setminus k}(F +\sum_{\nu\in{\cal M}(j)\setminus\mu}\widehat{x}_{\nu j}
)\tau_j \right]}{\mbox{ Tr}_{\{\tau_j\}} \exp\left[+\beta
{\cal J}_\mu \prod_{j\in{\cal L}(\mu)\setminus k}\tau_j +
\sum_{j\in{\cal L}(\mu)\setminus k}(F + \sum_{\nu\in{\cal M}(j)\setminus
\mu}\widehat{x}_{\nu j} )\tau_j \right]}.
\label{eq:fieldrecur}
\end{equation}
By computing the traces and taking
$\beta\rightarrow\infty$ one obtains:
\begin{equation}
 \widehat{x}_{\mu k}=\mbox{atanh}\left[{\cal J}_\mu\prod_{j\in{\cal
 L}(\mu)\setminus k}\mbox{tanh}(F+\sum_{\nu\in{\cal M}(j)\setminus\mu}\widehat{x}_{\nu j})\right]
\label{eq:fields}
\end{equation}
The effective local magnetisation due to interactions with the nearest
neighbours in one branch is given by $\widehat{m}_{\mu
j}=\mbox{tanh}(\widehat{x}_{\mu j})$. The effective local field on a
base spin $j$ of a polygon $\mu$ due to $C-1$ branches in the previous
generation and due to the external field is $x_{\mu
j}=F+\sum_{\nu\in{\cal M}(j)\setminus\mu}\widehat{x}_{\nu j} $;
 the effective local magnetisation is, therefore,
$m_{\mu j}=\mbox{tanh}(x_{\mu j})$. Equation (\ref{eq:fields}) can
then be rewritten in terms of $\widehat{m}_{\mu j}$ and $m_{\mu j}$
and the PP equations \cite{mackay,KS,kschischang} can be recovered:
\begin{equation}
m_{\mu k} =\mbox{tanh}\left(F+\sum_{\nu\in{\cal M}(j)\setminus
\mu}\mbox{atanh }(\widehat{m}_{\nu k}) \right) \qquad \widehat{m}_{\mu
k} = {\cal J}_\mu \prod_{j\in{\cal L}(\mu)\setminus k}m_{\mu j}
\label{eq:BP}
\end{equation}

Once the magnetisations on the boundary ($0$-th generation) are
assigned, the local magnetisation $m_j$ in the central site is determined by
 iterating (\ref{eq:BP}) and computing :
\begin{equation}
m_j =\mbox{tanh}\left(F+\sum_{\nu\in{\cal M}(j)}\mbox{atanh
}(\widehat{m}_{\nu j}) \right)
\label{eq:maglocal}
\end{equation}
The free energy can be obtained by integration as (\ref{eq:BP}) represents 
extrema  of a  free energy \cite{MKSV,VSK2,bowman}.

\begin{figure}
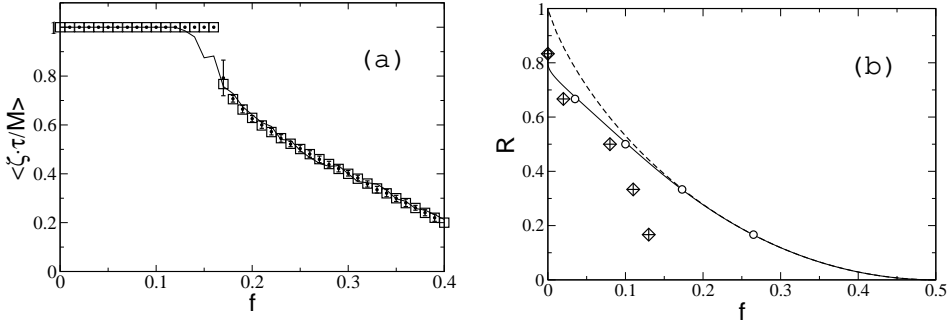

\begin{center}
\epsfig{file=overlap.eps,angle=-90,width=60mm}
\hspace{0.4cm}\epsfig{file=K6phase.eps,angle=-90,width=60mm}
\caption{(a) Mean normalised overlap between the actual noise vector
$\vetor{\zeta}$ and decoded noise $\widehat{\vetor{\tau}}$ for $K=4$
and $C=3$ (therefore $R=1/4$). Theoretical values ($\Box$),
experimental averages over $20$ runs for code word lengths $M=5000$
($\mbox{\small$\bullet$}$) and $M=100$ (full line). (b) Transitions
for $K=6$. Shannon's bound (dashed line), information theory upper
bound (full line) and thermodynamic transition obtained numerically
($\mbox{\small$\circ$}$). Theoretical ($\Diamond$) and experimental
($+$, $M=5000$ averaged over $20$ runs) PP decoding transitions are
also shown.  In both figures, symbols are chosen larger than the error bars.}
\label{fig2}
\end{center}
\end{figure}

By applying the gauge transformation ${\cal J}_\mu \mapsto 1$ and
$\tau_j \mapsto \tau_j\zeta_j$, assigning the probability distributions
$P_0(x)$ to boundary fields and averaging over random local fields
$F\zeta$ one obtains from (\ref{eq:fields}) the recursion relation 
in the space of probability distributions $P(x)$ \cite{bowman}:
\begin{eqnarray}
 P_n(x)&=&\int \prod_{l=1}^{C-1}d\widehat{x}_l\;
 \widehat{P}_{n-1}(\widehat{x}_l)\;\left\langle \delta\left[x- F\zeta
 -\sum_{l=1}^{C-1}\widehat{x}_l \right]\right\rangle_\zeta\nonumber \\
 \widehat{P}_{n-1}(\widehat{x})&=& \int \prod_{j=1}^{K-1}dx_j\;
 P_{n-1}(x_j) \;\delta\left[\widehat{x}- \mbox{atanh}
 \left(\prod_{j=1}^{K-1}\mbox{tanh}(x_j)\right)\right],
\label{eq:funfields}
\end{eqnarray}
where $P_n(x)$ is the distribution of effective fields at the $n$-th
generation due to the previous generations and external fields, in the
thermodynamic limit the distribution far from the boundary  will be 
$P_\infty(x)$ (generation $n\rightarrow\infty$). The local  field distribution at the central site  is computed by replacing $C-1$ by $C$ in (\ref{eq:funfields}), taking into account $C$ polygons in the generation just before the central site,  and inserting the  distribution $P_\infty(x)$. Equations (\ref{eq:funfields}) are identical to those obtained by the replica symmetric theory as in
\cite{KMS,MKSV,VSK2}.

By setting initial (boundary) conditions $P_0(x)$ and numerically
iterating (\ref{eq:funfields}), for $C\ge3$ one can find, up to some noise
 level $f_s$, a single stable fixed point at infinite fields, corresponding to
 a totally aligned state (successful decoding). At $f_s$ a bifurcation occurs 
and two other fixed points appear, stable and unstable, the former
corresponding to a misaligned state (decoding failure). This situation
is identical to that one observed in \cite{KMS,MKSV,VSK2}.  In terms
of the local fields distribution $P_n(x)$, the aligned state
corresponds to a runaway wave travelling to $x(n)\rightarrow\infty$ with $n$
being the time variable. The misaligned state corresponds to a stable
wave located at $x(n)\sim{\cal O}(1)$. Representing the
distributions (\ref{eq:funfields}) by the first cummulants only, one
can obtain a rough approximation in terms of one dimensional maps
showing a bifurcation at some noise level $\tilde{f}_s$, this approach will be
 further exploited elsewhere.

The practical PP decoding is performed by setting initial conditions
as $m_{\mu j}=1-2f$ to correspond to the prior probabilities and
iterating (\ref{eq:BP}) until stationarity or a maximum number of
iterations is attained \cite{mackay}. The estimate for the noise
vector is then produced by computing
$\widehat{\tau}_j=\mbox{sign}(m_j)$. At each decoding step the system  can
 be described by histograms of the variables (\ref{eq:BP}), this is equivalent
 to iterating (\ref{eq:funfields}) (a similar idea was presented in 
\cite{mackay,davey}).
 Below $f_s$ the process
always converges to the successful decoding state, above $f_s$ it
converges to the successful decoding only if the initial conditions
are fine tuned, in general the process converges to the failure
state. In Fig.\ref{fig2}a we show the theoretical mean overlap between
actual noise $\vetor{\zeta}$ and the estimate $\widehat{\vetor{\tau}}$ as a function of the noise level $f$ as
well as results obtained with PP decoding.

Information theory provides an upper bound for the maximum attainable
code rate  by equalising the maximal information contents of the syndrome
 vector $\vetor{z}$ and of the noise estimate
$\widehat{\vetor{\tau}}$ \cite{mackay,VSK2}. The thermodynamic phase
transition obtained by finding the stable fixed points of
(\ref{eq:funfields}) and their free energies interestingly coincides
with this upper bound within the precision of the numerical
calculation. Note that the performance predicted by thermodynamics is
not practical as it requires ${\cal O}(2^M)$ operations for an
exhaustive search for the global minimum of the free energy. In
Fig.\ref{fig2}b we show the thermodynamic  transition for $K=6$ and
compare with the upper bound, Shannon's bound and $f_s$ values.

\begin{table}
\caption{Gallager versus MN codes}
\[
\begin{tabular}{ccc}
\hline \mbox{ }&\mbox{ Gallager } &\mbox{ MN } \\ \hline
 \mbox{dynamical variables} &\mbox{\it M}& \mbox{\it N}+\mbox{\it M}\\ 
\mbox{constraints} &\mbox{\it M}-\mbox{\it N}&\mbox{\it M}\\ \mbox{unbiased messages
 coding}&\mbox{{\it for all  K}}&\mbox{\it K =1,2}\\ \mbox{Shannon's
 bound}&\mbox{{\it K}$\rightarrow\infty$}&\mbox{ {\it K}$\ge$ {\it 3 and unbiased messages}}\\ \hline
\end{tabular}
\]
\label{table}
\end{table}

The geometrical structure of a Gallager code defined by the matrix
$\vetor{A}$ can be represented by a bipartite graph ({\it Tanner
graph}) \cite{kschischang} with bit and check nodes. Each column $j$ of 
$\vetor{A}$ represents a bit node and each
row $\mu$ represents a check node, $A_{\mu j}=1$ means that there is
an edge linking bit $j$ to check $\mu$. It is possible to show
\cite{urbanke} that for a random ensemble of regular codes, the
probability of completing a cycle after walking $l$ edges starting
from an arbitrary node is upper bounded by ${\cal P}[l;K,C,M]\leq
l^2 K^{l}/M$. It implies that for  very large $M$ only cycles of 
at least order  $\ln M$ survive. In the thermodynamic limit
$M\rightarrow\infty$ the probability ${\cal P}[l;K,C,M]\rightarrow 0 $
for any finite $l$ and the bulk of the system is effectively
tree-like.  By mapping each check node to a polygon with $K$ bit nodes
as vertices, one can map a Tanner graph into a Husimi lattice that is
effectively a tree for any number of generations of order less than
$\ln M$. It is experimentally observed that the number of
iterations of (\ref{eq:BP}) required for convergence does not scale
with the system size, therefore, it is expected that the interior of a
tree-like lattice approximates a Gallager code with increasing
accuracy as the system size increases. Fig.\ref{fig2}a shows
that the approximation is fairly good even for sizes as small
as $M=100$. Note that although the local magnetisations $m_j$ for a
loopy graph are not generally expected to converge to the values
computed in a tree, $\mbox{sgn}(m_j)$ seems to do so. A thorough
discussion on this respect for some specific graphical models can be
found in \cite{weiss}.

In \cite{mackay} MacKay and Neal introduced a variation on Gallager
codes termed MN codes. The main difference between these codes is that
for MN codes the syndrome vector contains also information on the
original message in the form
$\vetor{z}=\vetor{C_s}\vetor{\xi}+\vetor{C_n}\vetor{\zeta}$. The
message itself is directly estimated and there is no need for
recovering the noise vector. MacKay has formulated and proved a number
of theorems simultaneously for both codes using the fact that if both
message and noise are sampled from the same distribution, these codes
can be formulated as the same estimation problem, to say, finding the
most probable vector $\vetor{x}$ that satisfies
$\vetor{z}=\vetor{A}\vetor{x}$, given the matrix $\vetor{A}$ and a
prior distribution $P(\vetor{x})$.  Using statistical physics, we
previously analysed MN codes \cite{KMS,MKSV,VSK2}. It is interesting
to note that in spite of the similarity between the two codes, there
are some important differences in their dependence on the parameters
$K$ and $C$. In particular, Shannon's bound is only attainable by
Gallager codes if $K\rightarrow\infty$, in contrast to results
obtained for MN codes. Decoding of unbiased messages is generally
possible with Gallager codes, but successful convergence is only
guaranteed (in the thermodynamic limit) for $K=1,2$ in the MN
codes. We outlined those differences in table \ref{table}.

To summarise, we solved exactly, without resorting to the replica
method, a system representing a Gallager  code on a Husimi cactus. 
 The results obtained are in agreement with the replica symmetric 
calculation  and with numerical experiments carried out in systems of 
moderate size. The framework can be  easily extended to MN and similar codes. 
We believe that methods of statistical physics are complimentary to those 
used in the statistical inference community and can enhance our understanding
 of general graphical models beyond error-correcting codes.

\stars

 We acknowledge support from EPSRC  (GR/N00562), The Royal Society (RV,DS)
and from the JSPS RFTF program (YK).

\end{document}